\documentstyle [12pt,a4p,epsfig,amsmath,multicol]{article}
\begin{document}
\vspace*{0.6cm}

\begin{center} 
{\normalsize\bf Proposal for a Satellite-Borne Experiment to Test Relativity
 of Simultaneity in Special Relativity}
\end{center}
\vspace*{0.6cm}
\centerline{\footnotesize J.H.Field}
\baselineskip=13pt
\centerline{\footnotesize\it D\'{e}partement de Physique Nucl\'{e}aire et 
 Corpusculaire, Universit\'{e} de Gen\`{e}ve}
\baselineskip=12pt
\centerline{\footnotesize\it 24, quai Ernest-Ansermet CH-1211Gen\`{e}ve 4. }
\centerline{\footnotesize E-mail: john.field@cern.ch}
\baselineskip=13pt
 
\vspace*{0.9cm}
\abstract{ An orbiting `photon clock' is proposed to test directly the
 relativity of simultaneity of special relativity. This is done by comparison
 of the arrival times at a ground station of three microwave signals
 transmitted by two satellites following the same low Earth orbit.}
\vspace*{0.9cm}
\normalsize\baselineskip=15pt
\setcounter{footnote}{0}
\renewcommand{\thefootnote}{\alph{footnote}}

 Einstein's original paper on special relativity (SR)~\cite{Ein1} discussed three
 physical effects which changed completely our conceptual understanding of space and time.
 These are: relativity of simultaneity (RS), length contraction (LC) and time dilatation (TD).
 RS and LC are closely related. In the space-time Lorentz transformation (LT)
 \footnote{The frame S' (space-time coordinates $x'$,$t'$) moves along the positive $x$-axis of
   S (space-time coordinates $x$,$t$) with velocity $v$. O$x'$ is parallel to O$x$. 
   Clocks in S and S' are synchronised so that $t= t' =0$ when the origins of S and S' coincide.
    $\beta \equiv v/c$, $\gamma \equiv 1/\sqrt{1-\beta^2}$.}:
   \begin{eqnarray}
   x' & = & \gamma(x-vt) \\
   t' & = & \gamma(t-\frac{\beta x}{c})
   \end{eqnarray}
   both effects result  from the spatial dependence (the term $-\gamma \beta x/c$) of the
   time transformation equation (2). LC is given by a $\Delta t = 0$ projection
   of the LT~\cite{JHF1}. Because of RS, events which are simultaneous in S (i.e.
   have $\Delta t = 0$) are not so in S', resulting in LC~\cite{JHF1,JHF2}. One 
   hundred years after the publication of Einstein's paper only the TD effect has been
   experimentally confirmed. For a concise review of experimental tests of SR see
   Reference~\cite{JHF3}. Unlike LC, the TD effect (a $\Delta x' = 0$ projection of
   the LT~\cite{JHF1}) does not involve RS since the space-time events concerned
   occur always at a fixed position in S' --the spatial coordinate of the clock
   under consideration. 
   \par The purpose of this letter is to propose a direct experimental test of RS.
    The enormous improvement, in recent decades, of the precision of
    time measurements due to the widespread application of atomic clocks much facilitates
   the test. A test of the related LC effect seems, in contrast, to be much more
   difficult~\cite{JHF3}. The proposed experiment is an actual realisation, in space, of
   the light-signal clock synchronisation procedure proposed by Einstein in ~\cite{Ein1}.
   However, no actual experimental clock synchronisation is needed. At the practical
   level the experiment can be considered as a sequel to the Spacelab experiment
   NAVEX~\cite{NAVEX} in which special and general relativity (TD and the gravitational
    red-shift) were tested by comparing a caesium clock in a space shuttle in a low,
   almost-circular, orbit around the Earth with a similar, synchronised, clock
   at a ground station. The experiment requires two satellites, one of which could
    conveniently be the International Space Station (ISS) which has orbit parameters
   similar to those of the NAVEX shuttle, the other a shuttle, or other satellite,
   following the same orbit as the ISS but separated from it by a few hundred 
   kilometers.
\begin{figure}[htbp]
\begin{center}\hspace*{-0.5cm}\mbox{
\epsfysize15.0cm\epsffile{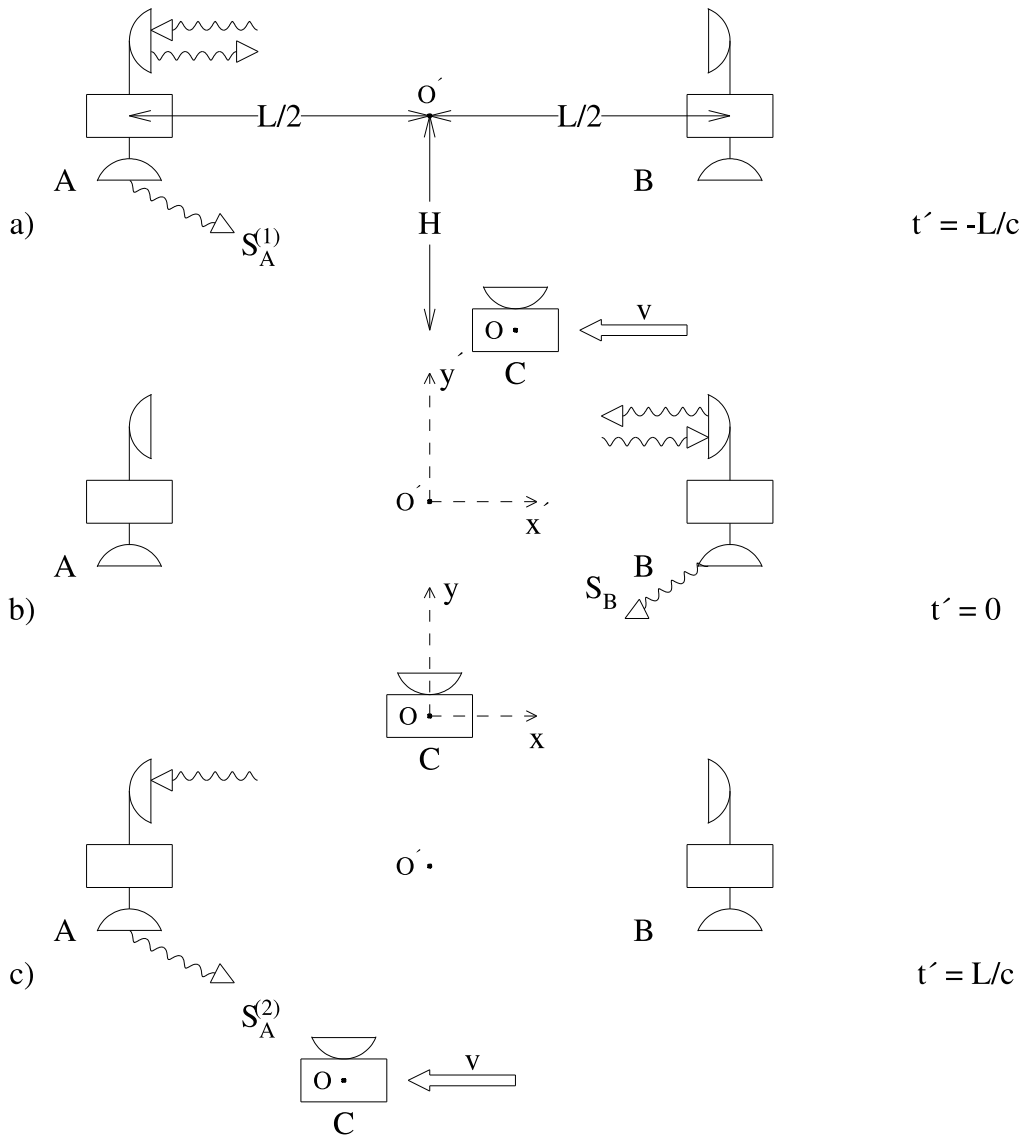}}
\caption{{\em Scheme of an experimental realisation of Einstein's clock synchronisation procedure
   using two satellites in low Earth orbit. The x-y projection is shown. 
  `Relativity of Simultaneity' is directly
 tested in the experiment  by observation at the ground station C of the times
  of arrival of the `photon clock' signals  $S_A^{(1)}$ and  $S_A^{(2)}$ from
  the satellite A [a) and c)] and $S_B$ from the satellite B [b)]. 
 C is viewed from the co-moving frame of A and B. Coordinate systems and geometrical
   and temporal prameters used in the analysis are defined.}}
\label{fig-fig1}
\end{center}
\end{figure}
   \par A scheme of the proposed experiment is shown in Fig.1. Two satellites, A and B, in
    low Earth orbit, separated by the distance $L$, pass near to a ground station C. Cartesian
    coordinate systems are defined in the co-moving inertial frame of the satellites (S')
   and the ground station (S). The origin of S' is chosen midway between A and B with
   $x'$-axis parallel to the direction of motion of the satellites and $y'$  axis outwardly
   directed in the plane of the orbit. O$x$ and O$y$ are parallel to  O'$x$ and O'$y$
   at the position of closest approach (culmination) of O' to C. Clocks in S and S' 
   are synchronised at $t = t' =0$ at culmination, where the coordinates of O' in S are:
   (x,y,z) = (0,$H$,$D$) and the relative velocity of S and S' is $v$. It is assumed in the
   following that for space time events at the satellites the LT equations (1) and (2)
   are valid between the frames S and S', not only at  $t = t' =0$, but for neighbouring
   times.
   \par A microwave signal is sent from B towards A so as to arrive there at the time 
   $t' = -L/c$ (Fig.1a). The signal is detected and reflected promptly back towards B.
   After a delay $t_D(A)$ the signal $S_A^{(1)}$ is sent from A to C. The reflected signal
    from A arrives back at B at time $t' =0$ (Fig.1b). It is detected and reflected
    promptly back towards A. After a delay  $t_D(B)$ the signal $S_B$ is sent from B to C.
    At time  $t' = L/c$ the inter-satellite signal arrives for a second time at A and
    after the delay  $t_D(A)$ sends the signal $S_A^{(2)}$ to C (Fig1c). The space-time
    coordinates of the emission events of the signals $S_A^{(1)}$, $S_B$ and $S_A^{(2)}$,
    as calculated using the LT (1) and (2) are presented in Table 1. Taking into
    account the propagation times of the signals from A and B to C the 
    following differences of arrival times of the signals at C are found:
   \begin{eqnarray}
   \delta t_{BA} \equiv t(S_B)-t(S_A^{(1)})& = & \frac{L}{c} + \frac{L}{c\beta}(d_B-d_A)
     +\frac{L^2}{2cR}(d_B+d_A) + \frac{\beta L}{c} -\frac{\beta L^2}{2cR} \\
    \delta t_{AB} \equiv t(S_A^{(2)})-t(S_B) & = & \frac{L}{c} - \frac{L}{c\beta}(d_B-d_A)
     -\frac{L^2}{2cR}(d_B+d_A) - \frac{\beta L}{c} -\frac{\beta L^2}{2cR}
  \end{eqnarray}
    where $R \equiv \sqrt{H^2+D^2 +L^2/4}$, $d_{A,B} \equiv v t_D(A,B)/L$ and
   only terms of O($\beta$) have been retained.
   Hence:
 \begin{equation}
  \Delta t \equiv   \delta t_{BA} -  \delta t_{AB}  = \frac{2\beta L}{c}
    +\frac{2 L}{c\beta}(d_B-d_A) +\frac{L^2}{cR}(d_B+d_A)
  \end{equation}
   It is interesting to note that RS, as manifested in the non-vanishing value of
   $\Delta t$ in (5) when $t_D(A)= t_D(B)= 0$, is an O($\beta$) effect, not an O($\beta^2$) one
   as for LC and TD.
   The term $2 \beta L/c$ in (5) originates from the second (spatially-dependent) term in
   the LT of time,(2), responsible for RS.
   The orbital velocity of the ISS is 7.674 km/s ($\beta = 2.56\times 10^{-5}$)
   ~\cite{ISS}. Since the ground station velocity in much less than this\footnote{For the NAVEX 
   experiment it was 0.311km/s.}, this is essentially the same as the relative velocity
   in (5). Choosing $L = 400$km (for the ISS $H \simeq 350$km~\cite{ISS}) and setting 
   $t_D(A)= t_D(B)= 0$ in (5) gives $\Delta t = 2 \beta L/c = 68.3$ns. Such a time difference
   is easily measurable with modern techniques. Signal arrival times in the NAVEX experiment
   were quoted with 1ns precision. The uncertainities in the clock rates for the relativity
   tests in NAVEX corresponded to an experimental time resolution of $\simeq 0.1$ns over one
   rotation period (1.6h) of the shuttle. The contribution of the last term on the right
   side of (5) is negligible. For $L/R =1$ and delays as long as 1$\mu$s it contributes
   only 0.05ns to $\Delta t$ for $\beta = 2.56\times 10^{-5}$.
   Thus $\Delta t$ is essentially independent of the distance between the satellites
   and the ground station at culmination. During the total transit time of the 
  microwave signals in the `photon clock' constituted by the satellites, they move in S
    only a distance $\simeq 2L\beta = 10.2$m. Different times of 
    emission of the signal sequence are easily taken into account by a suitable
   choice of the delay times $t_D(A,B)$.

 \begin{table}
\begin{center}
\begin{tabular}{|c||c|c|c|c|} \hline
 Event   & $x'$ & $t'$  & $x$ & $t$  \\
 \hline \hline
 & & & & \\
 $S_A^{(1)}$ emitted & $-\frac{L}{2}$ & $-\frac{L}{c}+t_D(A)$ &  $-\gamma L(\frac{1}{2}
  +\beta- \frac{v t_D(A)}{L})$ &  $-\frac{\gamma L}{c}(1+\frac{\beta}{2}- \frac{c t_D(A)}{L})$ \\
 & & & & \\
$S_B$ emitted & $\frac{L}{2}$ & $t_D(B)$ &  $\gamma L(\frac{1}{2}+
  \frac{v t_D(B)}{L})$ &  $\frac{\gamma L}{c}(\frac{\beta}{2}+\frac{c t_D(B)}{L})$ \\
 & & & & \\
 $S_A^{(2)}$ emitted & $-\frac{L}{2}$ & $\frac{L}{c}+t_D(A)$ &  $-\gamma L(\frac{1}{2}
  -\beta- \frac{v t_D(A)}{L})$ &  $\frac{\gamma L}{c}(1-\frac{\beta}{2}+\frac{c t_D(A)}{L})$ \\
 & & & & \\
 \hline
\end{tabular}
\caption[]{{\em Coordinates of space time events in S' and S. The origin
   of S' is midway between the satellites A and B. The origin
  of S is at C.}}      
\end{center}
\end{table}

   \par Although a particular coordinate system and clock synchronisation
    are used to calculate the entries of Table 1 and the time differences 
    $\delta t_{BA}$ and $\delta t_{AB}$ in (5), the quantity $\Delta t$ is independent 
    of this choice, so no clock synchronisation is required to measure $\Delta t$.
     If, however, pre-synchronised clocks are availabe in the satellites they may
    be used to generate the signal sequence: $S_A^{(1)}$, $S_B$, $S_A^{(2)}$, without 
    the necessity of `photon clock' signals between the satellites. The latter in
   fact may be considered to effect a real-time synchronisation of hypothetical
   clocks in A and B. In this case, a simpler direct measurement of RS is possible.
   Sending signals $S_A$ and $S_B$ at the same time in S', when O' is
   at culmination, the LT (1) and (2) predict that the signals will be observed at C with
   a time difference of $\gamma \beta L/c$, half the value of $\Delta t$ in 
   the photon clock experiment. 
    \par The ease of measurement of the O($\beta$) RS effect may be contrasted
   with the difficulty of measuring, in a similar experiment, the  O($\beta^2$)
   LC effect. Using the value  $\beta = 2.56\times 10^{-5}$ appropriate for the ISS,
  the apparent contraction of the distance between the satellites A and B,
  as viewed at some instant in S, of $(1-1/\gamma)L$, amounts to only 131$\mu$m for
   $L = 400$km. It is hard to concieve any experiment using currently known
   techniques with sufficiently good spatial resolution to measure such a tiny effect.
    \par In a recent paper by the present author~\cite{JHF3} it has been suggested, in 
   order to avoid certain casual paradoxes of SR, and to ensure translational
    invariance, that the origin of the frame S' in the LT (1) and (2) should be chosen
   to coincide with the position of the transformed event (a `local' LT).
   In this case it is predicted that $\Delta t = 0$ for  $t_D(A)= t_D(B)= 0$ in (5)
   and that signals emitted simultaneously in S' from A and B at culmination will be
   received at the same time in S at C.
 


\begin{thebibliography}{99}
\bibitem{Ein1} 
A.Einstein, Annalen der Physik {\bf17}, 891 (1905).
\bibitem{JHF1}
J.H.Field,  Am. J. Phys. {\bf 68}, 367 (2000).
\bibitem{JHF2}
J.H.Field, `On the Real and Apparent Positions of Moving Objects
 in Special Relativity: The Rockets-and-String and Pole-and-Barn
  Paradoxes Revisited and a New Paradox', arXiv pre-print: physics/0403094 
\bibitem{JHF3}
J.H.Field,  `The Local Space-Time Lorentz Transformation: a New Formulation
 of Special Relativity Compatible with Translational Invariance',
 arXiv pre-print: physics/0501043.  
 \bibitem{NAVEX}
 E.Sappl, Naturwissenschaften {\bf 77}, 325 (1990).
  \bibitem{ISS}
 http://www.heavens-above.com/orbitdisplay.asp?satid=25544. 
\end{thebibliography}
\end{document}